\documentclass[10pt,conference]{IEEEtran}

\usepackage{cite}

\ifCLASSINFOpdf
  \usepackage[pdftex]{graphicx}
\else
  \usepackage[dvips]{graphicx}
\fi

\usepackage{xcolor}
\definecolor{mycolor}{rgb}{1, 0, 0}
\usepackage{tikz}

\usepackage[normalem]{ulem}

\usepackage{amsmath}

\usepackage{algorithmic}

\usepackage{url}

\usepackage{hyperref}

\begin{document}
\title{Variational learning for \\ quantum artificial neural networks}

\author{\IEEEauthorblockN{Francesco Tacchino\IEEEauthorrefmark{1}\IEEEauthorrefmark{4}\IEEEauthorrefmark{5}, Stefano Mangini\IEEEauthorrefmark{2}\IEEEauthorrefmark{6}\IEEEauthorrefmark{5}, Panagiotis Kl. Barkoutsos\IEEEauthorrefmark{1},\\ Chiara Macchiavello\IEEEauthorrefmark{2}\IEEEauthorrefmark{6}\IEEEauthorrefmark{7}, Dario Gerace\IEEEauthorrefmark{2}, Ivano Tavernelli\IEEEauthorrefmark{1} and Daniele Bajoni\IEEEauthorrefmark{3}} 
\IEEEauthorblockA{\IEEEauthorrefmark{1}IBM Quantum, IBM Research -- Zurich, 8803 R\"{u}schlikon, Switzerland}
\IEEEauthorblockA{\IEEEauthorrefmark{2}University of Pavia,
Department of Physics, via Bassi 6, 27100 Pavia, Italy}
\IEEEauthorblockA{\IEEEauthorrefmark{3}University of Pavia,
Department of Industrial and Information Engineering,
via Ferrata 1, 27100 Pavia, Italy}
\IEEEauthorblockA{\IEEEauthorrefmark{6}INFN Sezione di Pavia, Via Bassi 6, I-27100, Pavia, Italy}
\IEEEauthorblockA{\IEEEauthorrefmark{7}CNR-INO - Largo E. Fermi 6, I-50125, Firenze, Italy}
\IEEEauthorblockA{\IEEEauthorrefmark{4}Email: fta@zurich.ibm.com}
\IEEEauthorblockA{\IEEEauthorrefmark{5}These authors contributed equally to this work.}}

\maketitle

\begin{abstract}

In the last few years, quantum computing and machine learning fostered rapid developments in their respective areas of application, introducing new perspectives on how information processing systems can be realized and programmed. The rapidly growing field of Quantum Machine Learning aims at bringing together these two ongoing revolutions. Here we first review a series of recent works describing the implementation of artificial neurons and feed-forward neural networks on quantum processors. We then present an original realization of efficient individual quantum nodes based on variational unsampling protocols. We investigate different learning strategies involving global and local layer-wise cost functions, and we assess their performances also in the presence of statistical measurement noise. While keeping full compatibility with the overall memory-efficient feed-forward architecture, our constructions effectively reduce the quantum circuit depth required to determine the activation probability of single neurons upon input of the relevant data-encoding quantum states. This suggests a viable approach towards the use of quantum neural networks for pattern classification on near-term quantum hardware.
\end{abstract}

\IEEEpeerreviewmaketitle

\section{Introduction}
\label{sec:introduction}
In classical machine learning, artificial neurons and neural networks were originally proposed, more than a half century ago, as trainable algorithms for classification and pattern recognition~\cite{McCulloch_Pitts_1943,rosenblatt_perceptron:_1957}. A few milestone results obtained in subsequent years, such as the backpropagation algorithm~\cite{rumelhart_learning_1986} and the Universal Approximation Theorem~\cite{cybenko_approximation_1989,hornik_approximation_1991}, certified the potential of deep feed-forward neural networks as a computational model which nowadays constitutes the cornerstone of many artificial intelligence protocols~\cite{Zurada:intro_ANN_1992,Rojas_ANN_Introduction}.

In recent years, several attempts were made to link these powerful but computationally intensive applications to the rapidly growing field of quantum computing, see also Ref.~\cite{biamonte_quantum_2017} for a useful review. The latter holds the promise to achieve relevant advantages with respect to classical machines already in the near term, at least on selected tasks including e.g.\ chemistry calculations~\cite{kandala_hardware-efficient_2017,cao_quantum_2019}, classification and optimization problems~\cite{moll_quantum_2018}. Among the most relevant results obtained in Quantum Machine Learning it is worth mentioning the use of trainable parametrized digital and continuous-variable quantum circuits as a model for quantum neural networks~\cite{farhi_classification_2018,grant_hierarchical_2018,mcclean_barren_2018,killoran_continuous-variable_2018,benedetti_parameterized_2019,mari_transfer_2019,cong_quantum_2019,PhysRevA.101.032308,henderson_quanvolutional_2020,broughton_tensorflow_2020}, the realization of quantum Support Vector Machines (qSVMs)~\cite{rebentrost_quantum_2014} working in quantum-enhanced feature spaces~\cite{havlicek_supervised_2019,schuld_quantum_2019} and the introduction of quantum versions of artificial neuron models~\cite{schuld_simulating_2015,wiebe_quantum_2016,cao_quantum_2017,tacchino_artificial_2019,torrontegui_unitary_2019,tacchino_quantum_2019,kristensen_artificial_2019,Mangini_2020}. However, it is true that very few clear statements have been made concerning the concrete and quantitative achievement of quantum advantage in machine learning applications, and many challenges still need to be addressed~\cite{biamonte_quantum_2017,wright_capacity_2019,chia_sampling-based_2019}.

In this work, we review a recently proposed quantum algorithm implementing the activity of binary-valued artificial neurons for classification purposes.  Although formally exact, this algorithm in general requires quite large circuit depth for the analysis of the input classical data. To mitigate for this effect we introduce a variational learning procedure, based on quantum unsampling techniques, aimed at critically reducing the quantum resources required for its realization. By combining memory-efficient encoding schemes and low-depth quantum circuits for the manipulation and analysis of quantum states, the proposed methods, currently at an early stage of investigation, suggest a practical route towards problem-specific instances of quantum computational advantage in machine learning applications.

\section{A model of quantum artificial neurons}
\label{sec:quantum_neuron}
The simplest formalization of an artificial neuron can be given following the classical model proposed by McCulloch and Pitts~\cite{McCulloch_Pitts_1943}. In this scheme, a single node receives a set of binary inputs $\{i_0,\ldots,i_{m-1}\} \in \{-1,1\}^m$, which can either be signals from other neurons in the network or external data. The computational operation carried out by the artificial neuron consists in first weighting each input by a synapse coefficient $w_{j} \in \{-1,1\}$ and then providing a binary output $O \in \{-1,1\}$ denoting either an active or rest state of the node determined by an integrate-and-fire response
\begin{equation}
O = \begin{cases} 1 \quad & \text{if } \sum_j w_{j}i_j \geq \theta \\
-1 \quad & \text{otherwise }
 \end{cases}
 \label{eq:McP-operation}
\end{equation}
where $\theta$ represents some predefined threshold.

A quantum procedure closely mimicking the functionality of a binary valued McCulloch-Pitts artificial neuron can be designed by exploiting, on one hand, the superposition of computational basis states in quantum registers, and on the other hand the natural non-linear activation behavior provided by quantum measurements. In this section, we will briefly outline a device-independent algorithmic procedure~\cite{tacchino_artificial_2019} designed to implement such a computational model on a gate-based quantum processor. More explicitly, we show how classical input and weight vectors of size $m$ can be encoded on a quantum hardware by using only $N = \log_2 m$ qubits
~\cite{schuld_implementing_2017,rebentrost_quantum_2018,tacchino_artificial_2019}. For loading and manipulation of data, we describe a protocol based on the generation of quantum hypergraph states~\cite{rossi_quantum_2013}. This exact approach to artificial neuron operations will be used in the main body of this work as a benchmark to assess the performances of approximate variational techniques designed to achieve more favorable scaling properties in the number of logical operations with respect to classical counterparts.

Let $\vec{i}$ and $\vec{w}$ be binary input and weight vectors of the form
\begin{equation}
\vec{i} = \begin{pmatrix}
    i_{0} \\
    i_{1} \\
    \vdots \\
    i_{m-1}
\end{pmatrix}\quad
\vec{w} = \begin{pmatrix}
    w_{0} \\
    w_{1} \\
    \vdots \\
    w_{m-1}
\end{pmatrix}
\label{eq:bin_vectors}
\end{equation}
with $i_j,w_j \in \{-1,1\}$ and $m = 2^{N}$. A simple and qubit-effective way of encoding such collections of classical data can be given by making use of the relative quantum phases (i.e.\ factors $\pm 1$ in our binary case) in equally weighted superpositions of computational basis states. We then  define the states
\begin{equation}
\begin{aligned}
|\psi_i\rangle = \frac{1}{\sqrt{m}}\sum_{j = 0}^{m - 1} i_j |j\rangle \\
|\psi_w\rangle = \frac{1}{\sqrt{m}}\sum_{j = 0}^{m - 1} w_j |j\rangle
\end{aligned}
\label{eq:inputstate}
\end{equation}
where, as usual, we label computational basis states with integers $j\in\{0,\ldots,m-1\}$ corresponding to the decimal representation of the respective binary string. The set of all possible states which can be expressed in the form above is known as the class of hypergraph states~\cite{rossi_quantum_2013}.

According to Eq.~\eqref{eq:McP-operation}, the quantum algorithm must first perform the inner product $\vec{i} \cdot \vec{w}$. It is not difficult to see that, under the encoding scheme of Eq.~\eqref{eq:inputstate}, the inner product between inputs and weights is contained in the overlap~\cite{tacchino_artificial_2019}
\begin{equation}
\langle \psi_w | \psi_i\rangle = \frac{\vec{w}\cdot\vec{i}}{m}
\end{equation}
We can explicitly compute such overlap on a quantum register through a sequence of $\vec{i}$- and $\vec{w}$-controlled unitary operations.  First, assuming that we operate on a N-qubit quantum register starting in the blank state $|0\rangle^{\otimes N}$, we can load the input-encoding quantum state $|\psi_i\rangle$ by performing a unitary transformation $\mathrm{U}_i$ such that
\begin{equation}
\mathrm{U}_i|0\rangle^{\otimes N}=|\psi_i\rangle
\end{equation}
It is important to mention that this preparation step would most effectively be replaced by, e.g., a direct call to a quantum memory~\cite{giovannetti_quantum_2008}, or with the supply of data encoding states readily generated in quantum form by quantum sensing devices to be analyzed or classified. It is indeed well known that the interface between classical data and their representation on quantum registers currently constitutes one of the major bottlenecks for Quantum Machine Learning applications~\cite{biamonte_quantum_2017}. Let now $\mathrm{U}_w$ be a unitary operator such that
\begin{equation}
\mathrm{U}_w |\psi_w\rangle = |1\rangle^{\otimes N} = |m-1\rangle
\label{eq:UwConstraint}
\end{equation}
In principle, any $m\times m$ unitary matrix having the elements of $\vec{w}$ appearing in the last row satisfies this condition. If we apply $\mathrm{U}_w$ after $\mathrm{U}_i$, the overall $N$-qubits quantum state becomes
\begin{equation}
\mathrm{U}_w |\psi_i\rangle = \sum_{j = 0}^{m - 1} c_j |j\rangle \equiv |\phi_{i,w}\rangle
\label{eq:afterUs}
\end{equation}
Using Eq.~\eqref{eq:UwConstraint}, we then have
\begin{equation}
\begin{aligned}
\langle \psi_w | \psi_i\rangle & = \langle \psi_w | U_w^\dagger U_w | \psi_i\rangle = \\
& = \langle m-1 |\phi_{i,w}\rangle = c_{m-1}
\end{aligned}
\label{eq:idotw}
\end{equation}
We thus see that, as a consequence of the constraints imposed to $\mathrm{U}_i$ and $\mathrm{U}_w$, the desired result $\vec{i} \cdot \vec{w} \propto \langle \psi_w | \psi_i\rangle$ is contained up to a normalization factor in the coefficient $c_{m-1}$ of the final state $|\phi_{i,w}\rangle$. 

The final step of the algorithm must access the computed input-weight scalar product and determine the activation state of the artificial neuron. In view of constructing a general architecture for feed-forward neural networks~\cite{tacchino_quantum_2019}, it is useful to introduce an ancilla qubit $a$, initially set in the state $|0\rangle$, on which the $c_{m-1} \propto \langle \psi_w | \psi_i\rangle$ coefficient can be written through a multi-controlled $\mathrm{NOT}$ gate, where the role of controls is assigned to the $N$ encoding qubits~\cite{tacchino_artificial_2019}:
\begin{equation}
|\phi_{i,w}\rangle|0\rangle_a \rightarrow \sum_{j = 0}^{m - 2} c_j |j\rangle|0\rangle_a + c_{m-1}|m-1\rangle|1\rangle_a
\label{eq:neuron-activation}
\end{equation}
At this stage, a measurement of qubit $a$ in the computational basis provides a probabilistic non-linear threshold activation behavior, producing the output $|1\rangle_a$ state, interpreted as an active state of the neuron, with probability $|c_{m-1}|^2$. Although this form of the activation function is already sufficient to carry out elementary classification tasks and to realize a logical $\mathrm{XOR}$ operation~\cite{tacchino_artificial_2019}, more complex threshold behaviors can in principle be engineered once the information about the inner product is stored on the ancilla~\cite{cao_quantum_2017,torrontegui_unitary_2019}. Equivalently, the ancilla can be used, via quantum controlled operations, to pass on the information to other quantum registers encoding successive layers in a feed-forward network architecture~\cite{tacchino_quantum_2019}. It is worth noticing that directing all the relevant information into the state of a single qubit, besides enabling effective quantum synapses, can be advantageous when implementing the procedure on real hardware on which readout errors constitute a major source of inaccuracy. Nevertheless, multi-controlled $\mathrm{NOT}$ operations, which are inherently non-local, can lead to complex decompositions into hardware-native gates especially in the presence of constraints in qubit-qubit connectivity. When operating a single node to carry out simple classification tasks or, as we will do in the following sections, to assess the performances of individual portions of the proposed algorithm, the activation probability of the artificial neuron can then equivalently be extracted directly from the register of $N$ encoding qubits by performing a direct measurement of $|\phi_{i,w}\rangle$ targeting the $|m-1\rangle \equiv |1\rangle^{\otimes N}$ computational basis state.

\subsection{Exact implementation with quantum hypergraph states}

A general and exact realization of the unitary transformations $\mathrm{U}_i$ and $\mathrm{U}_w$ can be designed by using the generation algorithm for quantum hypergraph states~\cite{tacchino_artificial_2019}. The latter have been extensively studied as useful quantum resources~\cite{rossi_quantum_2013,ghio_multipartite_2018}, and are formally defined as follows. Given a collection of $N$ vertices $V$, we call a $k$-hyper-edge any subset of exactly $k$ vertices. A hypergraph $g_{\leq N} = \{V,E\}$ is then composed of a set $V$ of vertices together with a set $E$ of hyper-edges of any order $k$, not necessarily uniform. Notice that this definition includes the usual notion of a mathematical graph if $k = 2$ for all (hyper)-edges. To any hypergraph $g_{\leq N}$ we associate a $N$-qubit quantum hypergraph state via the definition
\begin{equation}
|g_{\leq N}\rangle = \prod_{k=1}^N \prod_{\{q_{v_1},\dots,q_{v_k}\}\in E} \mathrm{C}^k \mathrm{Z}_{q_{v_1},\dots,q_{v_k}}|+\rangle^{\otimes N}
\label{eq:hyprgraphGeneral}
\end{equation}
where $q_{v_1},\dots,q_{v_k}$ are the qubits connected by a $k$-hyper-edge in $E$ and, with a little abuse of notation, we assume $\mathrm{C}^2 \mathrm{Z}\equiv\mathrm{CZ}$ and $\mathrm{C}^1 \mathrm{Z}\equiv \mathrm{Z} = \mathrm{R}_z(\pi)$. For $N$ qubits there are exactly $\mathcal{N} = 2^{2^N-1}$ different hypergraph states. We can make use of well known preparation strategies for hypergraph states to realize the unitaries $\mathrm{U}_i$ and $\mathrm{U}_w$ with at most a single $N$-controlled $\mathrm{C}^N\mathrm{Z}$ and a collection of $p$-controlled $\mathrm{C}^p\mathrm{Z}$ gates with $p<N$. It is worth pointing out already here that such an approach, while optimizing the number of multi-qubit logic gates to be employed, implies a circuit depth which scales linearly in the size of the classical input, i.e.\ $\mathcal{O}(m) \equiv \mathcal{O}(2^N)$, in the worst case corresponding to a fully connected hypergraph~\cite{tacchino_artificial_2019}.

To describe a possible implementation of $\mathrm{U}_i$, assume once again that the quantum register of $N$ encoding qubits is initially in the blank state $|0\rangle^{\otimes N}$. By applying parallel Hadamard gates ($\mathrm{H}^{\otimes N}$) we obtain the state $|+\rangle^{\otimes N}$, corresponding to a hypergraph with no edges. We can then use the target collection of classical inputs $\vec{i}$ as a control for the following iterative procedure:
\begin{algorithmic}
\FOR{$P=1$ to $N$}
{\FOR{$j=0$ to $m-1$}
{\IF{ ($|j\rangle$ has exactly $P$ qubits in $|1\rangle$ \AND $i_j = -1$)} 
\STATE Apply $\mathrm{C}^P \mathrm{Z}$ to those qubits
\STATE Flip the sign of $i_k$ in $\vec{i}$\ \ $\forall k$ such that $|k\rangle$ has the same $P$ qubits in $|1\rangle$
\ENDIF}
\ENDFOR}
\ENDFOR 
\end{algorithmic}
Similarly, $\mathrm{U}_w$ can be obtained by first performing the routine outlined obove (without the initial parallel Hadamard gates) tailored according to the classical control $\vec{w}$: since all the gates involved in the construction are the inverse of themselves and commute with each other, this step produces a unitary transformation bringing $|\psi_w\rangle$ back to $|+\rangle^{\otimes N}$. The desired transformation $\mathrm{U}_w$ is completed by adding parallel $\mathrm{H}^{\otimes N}$ and $\mathrm{NOT}^{\otimes N}$ gates~\cite{tacchino_artificial_2019}. 

\section{Variational realization of a quantum artificial neuron}

Although the implementation of the unitary transformations $\mathrm{U}_i$ and $\mathrm{U}_w$ outlined above is formally exact and optimizes the number of multi-qubit operations to be performed by leveraging on the correlations between the $\pm 1$ phase factors, the overall requirements in terms of circuit depth pose in general severe limitations to their applicability in non error-corrected quantum devices. Moreover, although with such an approach the encoding and manipulation of classical data is performed in an efficient way with respect to memory resources, the computational cost needed to control the execution of the unitary transformations and to actually perform the sequences of quantum logic gates remains bounded by the corresponding classical limits. Therefore, the aim of this section is to explore conditions under which some of the operations introduced in our quantum model of artificial neurons can be obtained in more efficient ways by exploiting the natural capabilities of quantum processors.

In the following, we will mostly concentrate on the task of realizing approximate versions of the weight unitary $\mathrm{U}_w$ with significantly lower implementation requirements in terms of circuit depth. Although most of the techniques that we will introduce below could in principle work equally well for the preparation of encoding states $|\psi_i\rangle$, it is important to stress already at this stage that such approaches cannot be interpreted as a way of solving the long standing issue represented by the loading of classical data into a quantum register. Instead, they are pursued here as an efficient way of \textit{analyzing} classical or quantum data presented in the form of a quantum state. Indeed, the variational approach proposed here requires ad-hoc training for every choice of the target vector $\vec{w}$ whose $\mathrm{U}_w$ needs to be realized. To this purpose, we require access to many copies of the desired $|\psi_w\rangle$ state, essentially representing a quantum training set for our artificial neuron. As in our formulation a single node characterized by weight connections $\vec{w}$ can be used as an elementary classifier recognizing input data sufficiently close to $\vec{w}$ itself~\cite{tacchino_artificial_2019}, the variational procedure presented here essentially serves the double purpose of training the classifier upon input of positive examples $|\psi_w\rangle$ and of finding an efficient quantum realization of such state analyzer.

\subsection{Global variational training}

\begin{figure*}[!t]
\centering
\begin{tikzpicture}
\node[inner sep=0pt] (russell) at (0,0){\includegraphics[width=2\columnwidth]{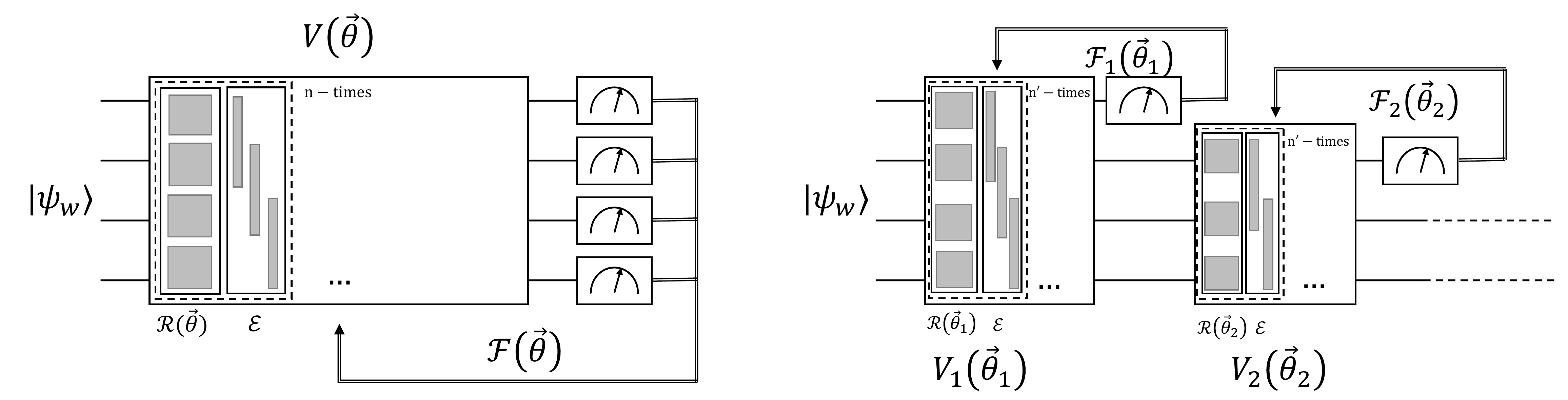}};
\node[inner sep=0pt] (russell) at (-9,1.8) {\normalsize{\textbf{a}}};
\node[inner sep=0pt] (russell) at (0,1.8) {\normalsize{\textbf{b}}};
\end{tikzpicture}
\caption{Variational learning via unsampling. (a) Global strategy, with optimization targeting  all qubits simultaneously. (b) Local qubit-by-qubit approach, in which each layer is used to optimize the operation for one qubit at a time.}
\label{fig:unsampling}
\end{figure*}

According to Eq.~\eqref{eq:UwConstraint}, the purpose of the transformation $\mathrm{U}_w$ within the quantum artificial neuron implementation is essentially to reverse the preparation of a non-trivial quantum state $|\psi_w\rangle$ back to the relatively simple product state $|1\rangle^{\otimes N}$. Notice that in general the qubits in the state $|\psi_w\rangle$ share multipartite entanglement~\cite{ghio_multipartite_2018}. Here we discuss a promising strategy for the efficient approximation of the desired transformation satisfying the necessary constraints based on variational techniques. Inspired by the well known variational quantum eigensolver (VQE) algorithm~\cite{peruzzo_variational_2014}, and in line with a recently introduced unsampling protocol~\cite{carolan_variational_2020}, we define the following optimization problem: given access to independent copies of $|\psi_w\rangle$ and to a variational quantum circuit, characterized by a unitary operation $ V(\vec{\theta})$ and parametrized by a set of angles $\vec{\theta}$, we wish to find a set of values $\vec{\theta}_{opt}$ that guarantees a good approximation of $\mathrm{U}_w$. The heuristic circuit implementation typically consists of sequential blocks of single-qubit rotations followed by entangling gates, repeated up to a certain number that guarantees enough freedom for the convergence to the desired unitary~\cite{kandala_hardware-efficient_2017}.

Once the solution $V(\vec{\theta}_{opt})$ is found, which in our setup corresponds to a fully trained artificial neuron, it would then provide a form of quantum advantage in the analysis of arbitrary input states $|\psi_i\rangle$ as long as the circuit depth for the implementation of the variational ansatz is sub-linear in the dimension of the classical data, i.e.\ sub-exponential in the size of the qubit register.
As it is customarily done in near-term VQE applications, the optimization landscape is explored by combining executions of quantum circuits with classical feedback mechanisms for the update of the $\vec{\theta}$ angles. 
In the most general scenario, and according to Eq.~\eqref{eq:UwConstraint}, a cost function can be defined as
\begin{equation}
    \mathcal{F}(\vec{\theta}) = 1 - |\langle 11\ldots 1| V(\vec{\theta})|\psi_w\rangle|^2
    \label{eq:global_cost_f}
\end{equation}
The solution $\vec{\theta}_{opt}$ is then represented by
\begin{equation}
    \vec{\theta}_{opt} = \arg \min_{\vec{\theta}} \mathcal{F}(\vec{\theta})
\end{equation}
and leads to $V(\vec{\theta}_{opt}) \simeq \mathrm{U}_w$. We call this approach a \textit{global} variational unsampling as the cost function in Eq.~\eqref{eq:global_cost_f} requires all qubits to be simultaneously found as close as possible to their respective target state $|1\rangle$, without making explicit use of the product structure of the desired output state $|1\rangle^{\otimes N}$. It is indeed well known that VQE can lead in general to exponentially difficult optimization problems~\cite{mcclean_barren_2018}, however the characteristic feature of the problem under evaluation may actually allow for a less complex implementation of the VQE for unsampling purposes~\cite{carolan_variational_2020}, as outlined in the following section.
A schematic representation of the global variational training is provided in Fig.~\ref{fig:unsampling}a. 

\subsection{Local variational training}

An alternative approach to the global unsampling task, particularly suited for the case we are considering in which the desired final state of the quantum register is fully unentangled, makes use of a local, qubit-by-qubit procedure. This technique, which was recently proposed and tested on a photonic platform as a route towards efficient certification of quantum processors~\cite{carolan_variational_2020}, is highlighted here as an additional useful tool within a general quantum machine learning setting.

In the local variational unsampling scheme, the global transformation $V(\vec{\theta})$ is divided into successive layers $V_j(\vec{\theta}_j)$ of decreasing complexity and size. Each layer is trained separately, in a serial fashion, according to a cost function which only involves the fidelity of a single qubit to its desired final state. More explicitly, every $V_j(\vec{\theta}_j)$ operates on qubits ${j,\ldots,N}$ and has an associated cost function
\begin{equation}
    \mathcal{F}_j(\vec{\theta}_j) = 1 - \langle 1 | \operatorname{Tr}_{j+1,\ldots,N}[\rho_j]|1\rangle
    \label{eq:local_cost_f}
\end{equation}
where the partial trace leaves only the degrees of freedom associated to the $j$-th qubit and, recursively, we define
\begin{equation}
    \rho_j = \begin{cases}
    V_j(\vec{\theta}_j)\rho_{j-1} V^\dagger_j(\vec{\theta}_j) \quad & j > 1 \\
    |\psi_w\rangle\langle\psi_w| \quad & j = 0
    \end{cases}
\end{equation}
At step $j$, it is implicitly assumed that all the parameters $\vec{\theta}_{k}$ for $k = 1,\ldots,j-1$ are fixed to the optimal values obtained by the minimization of the cost functions in the previous steps. Notice that, operationally, the evaluation of the cost function $\mathcal{F}_j$ can be automatically carried out by measuring the $j$-th qubit in the computational basis while ignoring the rest of the quantum register, as shown in Fig.~\ref{fig:unsampling}b.

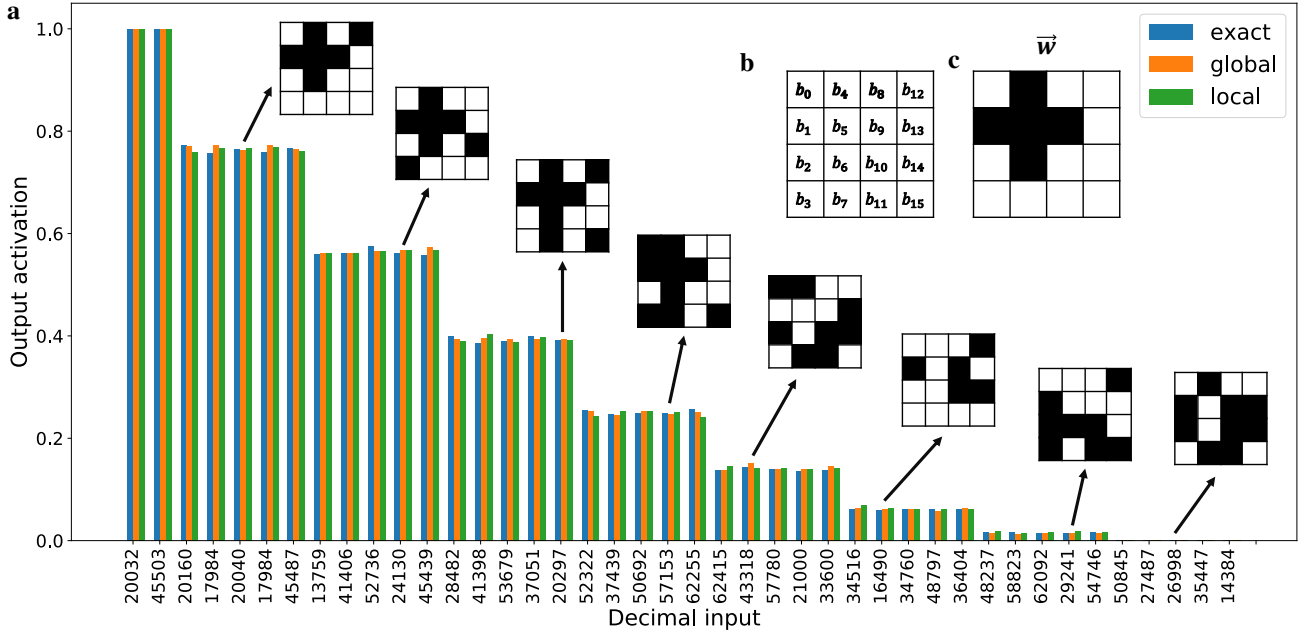
\begin{figure*}[!t]
\centering
\begin{tikzpicture}
\node[inner sep=0pt] (russell) at (0,0){\includegraphics[width=2\columnwidth]{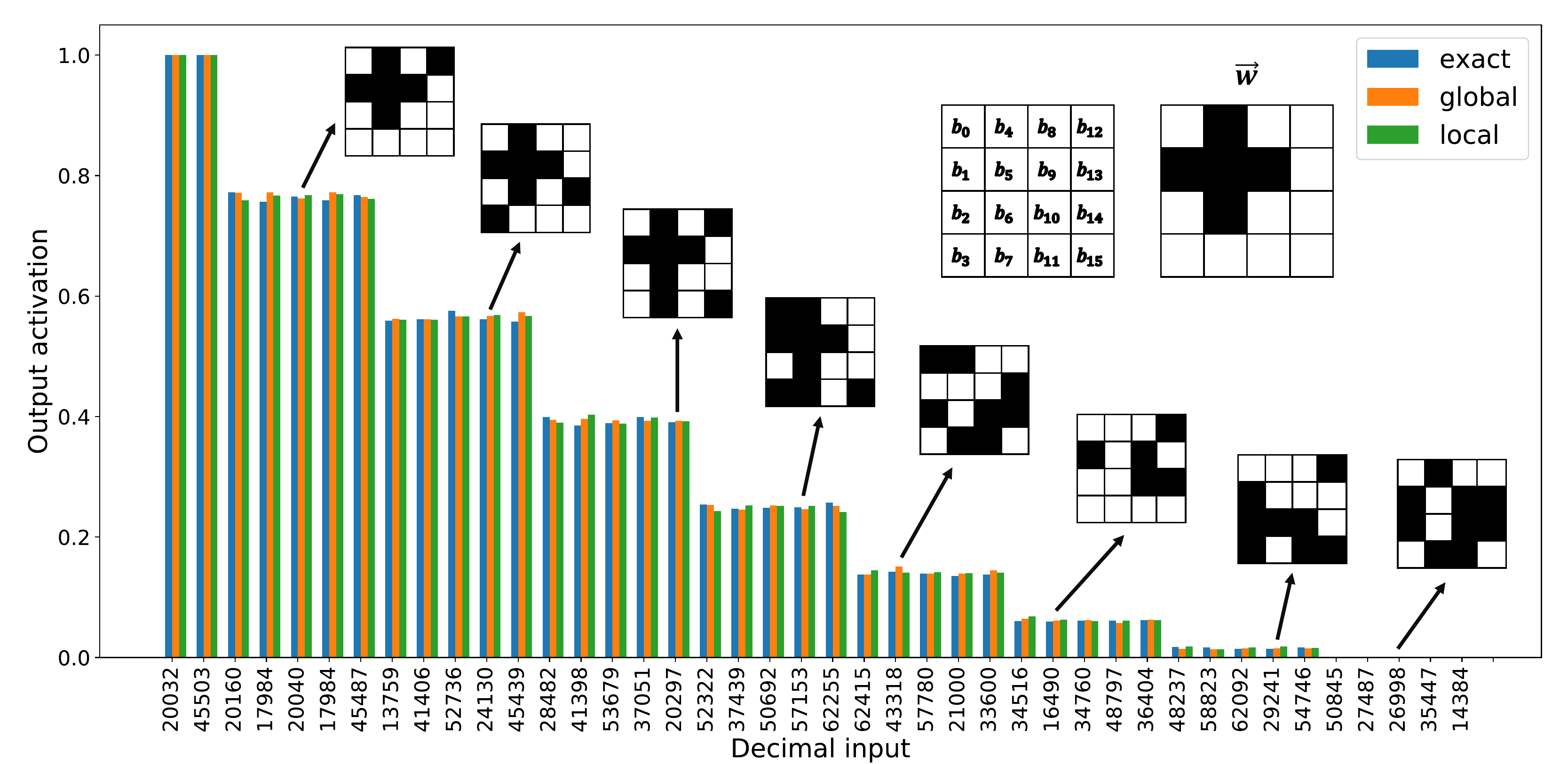}};
\node[inner sep=0pt] (russell) at (-8.5,3.9) {\normalsize{\textbf{a}}};
\node[inner sep=0pt] (russell) at (1.25,3.25) {\normalsize{\textbf{b}}};
\node[inner sep=0pt] (russell) at (4,3.25) {\normalsize{\textbf{c}}};
\end{tikzpicture}
\caption{Comparison of output activation $p_{out} = |\langle\psi_w|\psi_i\rangle|^2$ among the exact (hypergraph states routine), global ($n =3$) and local ($n'=2$) approximate implementations of $\mathrm{U}_w$. The inset shows the general mapping of any $16$-dimensional binary vector $\vec{b}$ onto the $4\times 4$ binary image (b) and the cross-shaped $\vec{w}$ used in this example (c). The selected inputs on which the approximations are tested were chosen to cover all the possible cases for $p_{out}$, and are labeled with their corresponding integer $k_i$ (see main text).}
\label{fig:test_inputs}
\end{figure*}

The benefits of local variational unsampling with respect to the global strategy are mainly associated to the reduced complexity of the optimization landscape per step. Indeed, the local version always operates on the overlap between single-qubit states, at the relatively modest cost of adding $N-1$ smaller and smaller variational ansatzes. In the specific problem at study, we thus envision the local approach to become particularly effective, and more advantageous than the global one, in the limit of large enough number of qubits, i.e.\ for the most interesting regime where the size of the quantum register, and therefore of the quantum computation, exceeds the current classical simulation capabilities.

\subsection{Case study: pattern recognition}
\label{sec:all2all}

To show an explicit example of the proposed construction, let us fix $m = 16$, $N = 4$. Following Ref.~\cite{tacchino_artificial_2019}, we can visualize a 16-bit binary vector $\vec{b}$, see Eq.~\eqref{eq:bin_vectors}, as a $4\times 4$ binary pattern of black ($b_j = -1$) and white ($b_j = 1$) pixels. Moreover, we can assign to every possible pattern an integer label $k_b$ corresponding to the conversion of the binary string $\mathtt{k}_b = \mathtt{b}_{0}\ldots \mathtt{b}_{15}$, where $b_j = (-1)^{\mathtt{b}_j}$. We choose as our target $\vec{w}$ the vector corresponding to $k_w = 20032$, which represents a black cross on white background at the north-west corner of the 16-bit image, see Fig~\ref{fig:test_inputs}.

Starting from the global variational strategy, we construct a parametrized ansatz for $V(\vec{\theta})$ as a series of entangling ($\mathcal{E}$) and rotation ($\mathcal{R}(\vec{\theta})$) cycles:
\begin{equation}
    V(\vec{\theta}) = \left(\prod_{c=1}^n \mathcal{R}(\theta_{c,1}\ldots\theta_{c,4})\mathcal{E}\right)\mathcal{R}(\theta_{0,1}\ldots\theta_{0,4})
\end{equation}
where $n$ is the total number of cycles which in principle can be varied to increase the expressibility of the ansatz by increasing its total depth. Rotations are assumed to be acting independently on the $N = 4$ qubits according to
\begin{equation}
    \mathcal{R}(\theta_{c,1}\ldots\theta_{c,4}) = \bigotimes_{q=1}^4\exp \left(-i\frac{\theta_{c,q}}{2}\sigma_y^{(q)}\right)
\end{equation}
where $\sigma_y^{(q)}$ is the Pauli $y$-matrix acting on qubit $q$. At the same time, the entangling parts promote all-to-all interactions between the qubits according to
\begin{equation}
    \mathcal{E} = \prod_{q}\prod_{q' = q+1}^4 \mathrm{CNOT}_{qq'}
    \label{eq:a2a_entangler}
\end{equation}
where $\mathrm{CNOT}_{qq'}$ is the usual controlled $\mathrm{NOT}$ operation between control qubit $q$ and target $q'$ acting on the space of all 4-qubits. For $n$ cycles, the total number $n_\theta$ of $\theta$-parameters, including the initial rotation layer $\mathcal{R}(\theta_{0,1}\ldots\theta_{0,4})$, is therefore $n_\theta = 4 + 4n$. 

A qubit-by-qubit version of the ansatz can be constructed in a similar way by using the same structure of entangling and rotation cycles, decreasing the total number of qubits by one after each layer of the optimization. Here we choose a uniform number $n'$ of cycles per qubit (this condition will be relaxed afterwards, see Sec.~\ref{sec:new_part}), thus setting $\forall j\neq 4$
\begin{equation}
    V_j(\vec{\theta}_j) = \left(\prod_{c=1}^{n'} \mathcal{R}(\theta_{c,j}\ldots\theta_{c,4})\mathcal{E}\right)\mathcal{R}(\theta_{0,j}\ldots\theta_{0,4})
    \label{eq:local_unitary}
\end{equation}
For $j = 4$, we add a single general single-qubit rotation with three parameters
\begin{equation}
    \label{eq:u3}
    V_4(\alpha,\beta,\gamma) =  \exp \left(-i\frac{(\alpha,\beta,\gamma)\cdot\vec{\sigma}^{(4)}}{2}\right)
\end{equation}
where $\vec{\sigma} = (\sigma_x,\sigma_y,\sigma_z)$ are again the usual Pauli matrices.

We implemented both versions of the variational training in Qiskit~\cite{Qiskit}, combining exact simulation of the quantum circuits required to evaluate the cost function with classical Nelder-Mead~\cite{10.1093/comjnl/7.4.308} and Cobyla~\cite{cobyla} optimizers from the scipy Python library. We find that the values $n = 3$ and $n'=2$ allow the routine to reach total fidelities to the target state $|1\rangle^{\otimes N}$ well above $99.99\%$. As shown in Fig~\ref{fig:test_inputs}, this in turn guarantees a correct reproduction of the exact activation probabilities of the quantum artificial neuron with a quantum circuit depth of $19$ $(29)$ for the global (qubit-by-qubit) strategy, as compared to the total depth equal to $49$ for the exact implementation of $\mathrm{U}_w$ using hypergraph states. This counting does not include the gate operations required to prepare the input state, i.e.\ it only evidences the different realizations of the $\mathrm{U}_w$ implementation assuming that each $|\psi_i\rangle$ is provided already in the form of a wavefunction. Moreover, the multi-controlled $\mathrm{C}^P\mathrm{Z}$ operations appearing in the exact version were decomposed into single-qubit rotations and $\mathrm{CNOT}$s without the use of additional working qubits. Notice that these conditions are the ones usually met in real near-term superconducting hardware endowed with a fixed set of universal operations.

\subsection{Structure of the ansatz and scaling properties}
\label{sec:new_part}
In many practical applications, the implementation of the entangling block $\mathcal{E}$ could prove technically challenging, in particular for near term quantum devices based, e.g., on superconducting wiring technology, for which the available connectivity between qubits is limited. For this reason, it is useful to consider a more hardware-friendly entangling scheme, which we refer to as \textit{nearest neighbours}. In this case, each qubit is entangled only with at most two other qubits, essentially assuming the topology of a linear chain
\begin{equation}
    \mathcal{E}_{\text{nn}} = \prod_{q=1}^3 \text{CNOT}_{q,q+1}
    \label{eq:nn_entangler}
\end{equation}
This scheme may require even fewer two-qubit gates to be implemented with respect to the all-to-all scheme presented above. Moreover, this entangling unitary fits perfectly well on those quantum processors consisting of linear chains of qubits or heavy hexagonal layouts.

We implemented both global and local variational learning procedures with nearest neighbours entanglers in Qiskit~\cite{Qiskit}, using exact simulation of the quantum circuits with classical optimizers to drive the learning procedure. In the following, we report an extensive analysis of the performances and a comparison with the all-to-all strategy introduced in Sec.~\ref{sec:all2all} above. All the simulations are performed by assuming the same cross-shaped target weight vector $\vec{w}$ depicted in Fig.~\ref{fig:test_inputs}.

\begin{figure}
    \centering
    \includegraphics[width=\columnwidth]{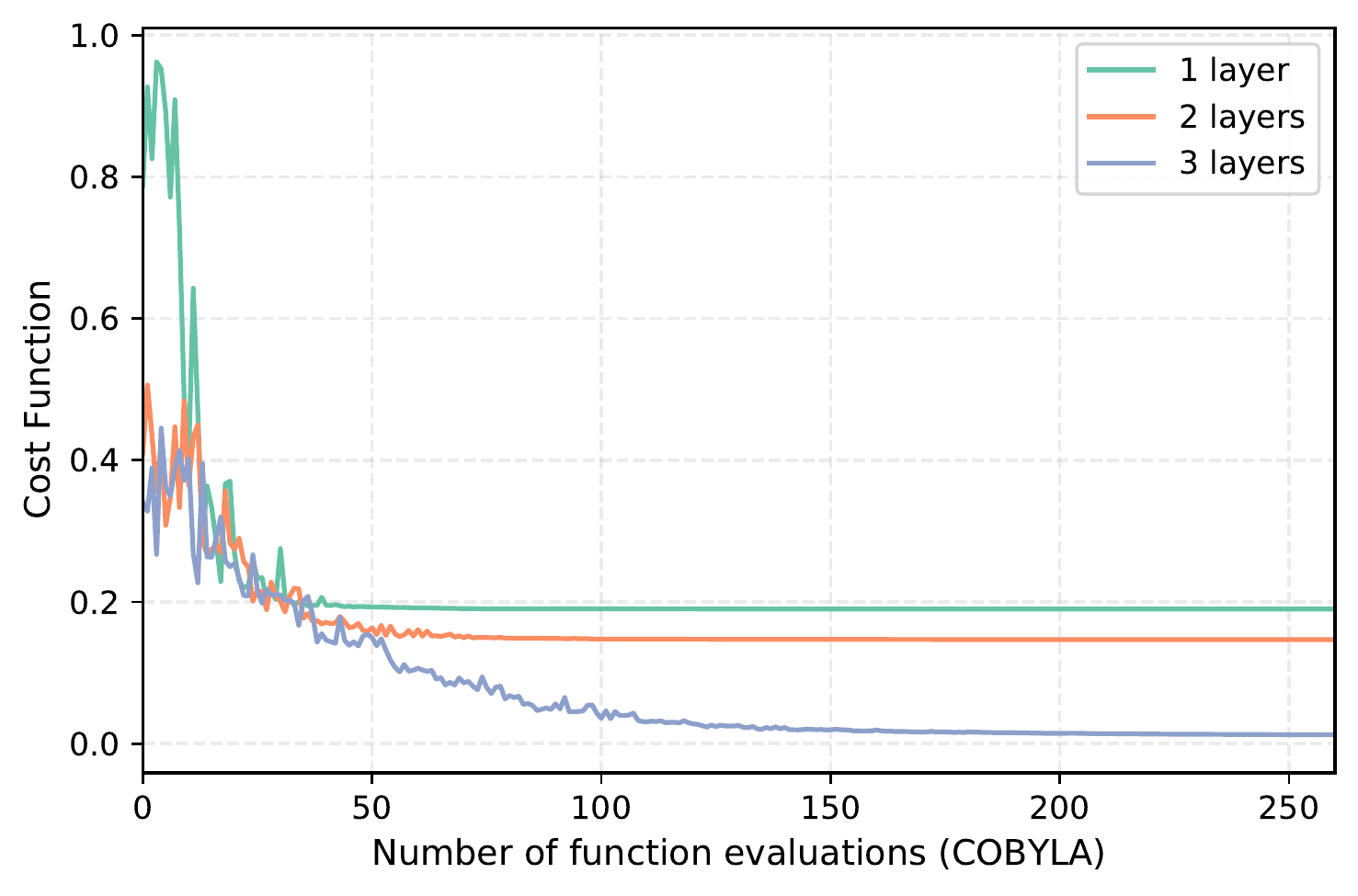}    
    \caption{Optimization of the global unitary with nearest neighbours entanglement for three different structures differing in the numbers of entangling blocks $n$. The cost function is $|\langle{11\ldots 1|V(\vec{\theta})|\psi_w\rangle}|^2 = 1 - \mathcal{F}(\vec{\theta})$, see Eq.~\eqref{eq:global_cost_f}. Only for $n=3$ the learning model has enough expressibility to reach a good final fidelity. The classical optimizer used in this case was COBYLA~\cite{cobyla}.}
    \label{fig:global_free}
\end{figure}

In Figure~\ref{fig:global_free} we show an example of the typical optimization procedure for three different choices of the ansatz depth (i.e.\ number of entangling cycles) $n=1,2,3$, assuming a global cost function. Here we find that $n=3$ allows the routine to reach a fidelity $\mathcal{F}(\vec{\theta})$ to the target state $|1\rangle^{\otimes N}$ above $99\%$. 

\begin{figure}
    \centering
    \includegraphics[width=\columnwidth]{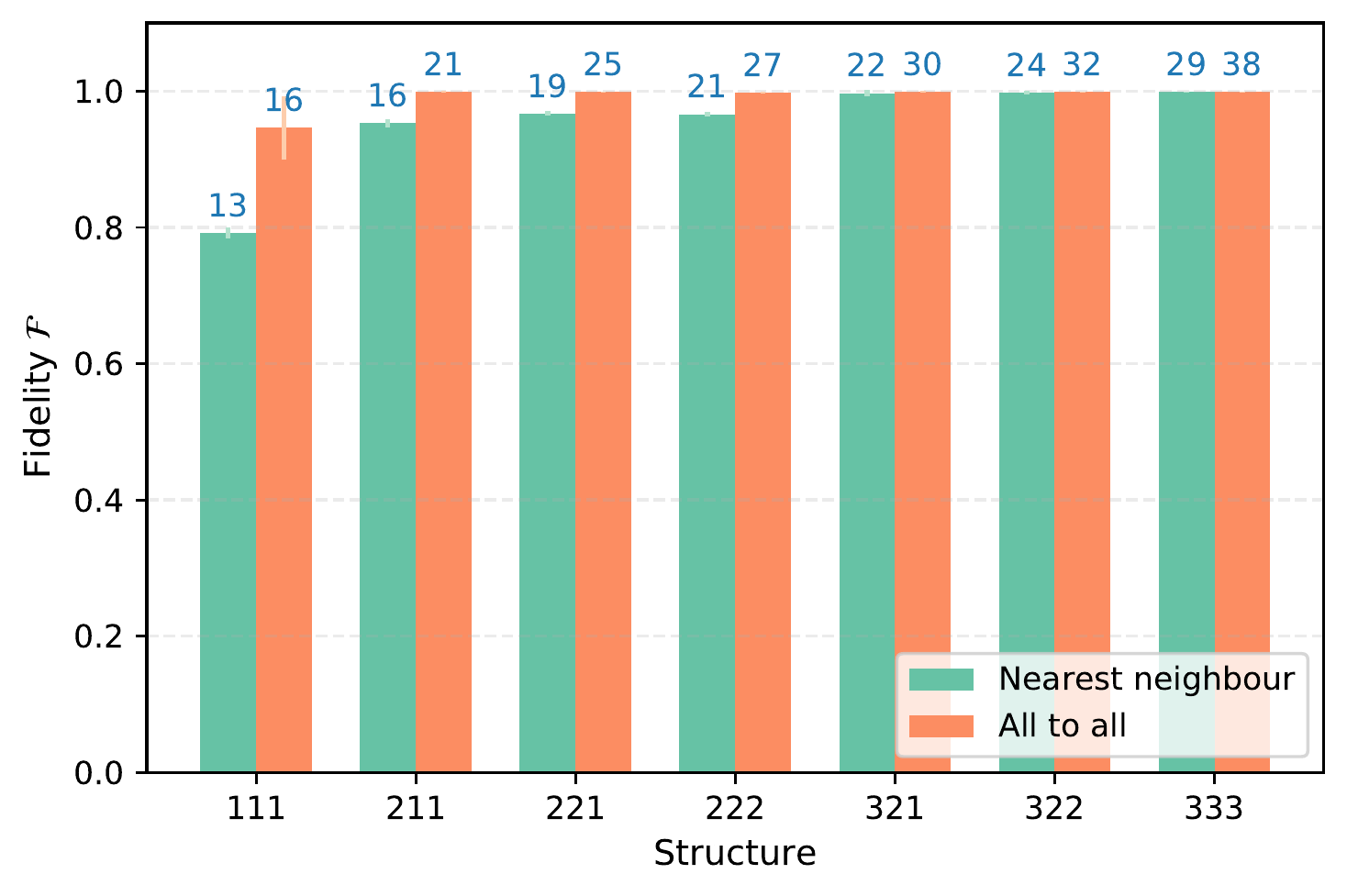}
    \caption{Final fidelity obtained for the local variational training and using both the all-to-all entangler $\mathcal{E}$~\eqref{eq:a2a_entangler} and nearest neighbour $\mathcal{E}_{\text{nn}}$~\eqref{eq:nn_entangler}. On top of each rectangle, in light blue, it is reported the depth of the corresponding quantum circuit to implement that given structure with that particular entangling scheme. For clarity, a structure `$211$' corresponds  to a variational model having two repetitions ($n'_1 = 2$) for the first layer acting on all 4 qubits, and 1 cycle ($n'_1=n'_2=1$) for the remaining two layers acting on $3$ and $2$ qubits respectively. Each bar was obtained executing the optimization process 10 times, and then evaluating the means and standard deviations (shown as error bars). The optimization procedure was performed using COBYLA~\cite{cobyla}.}
    \label{fig:nn_a2a_entangler}
\end{figure}

In the local qubit-by-qubit variational scheme, we can actually introduce an  additional degree of freedom by allowing the number of cycles per qubit, $n'$, to vary between successive layers corresponding to the different stages of the optimization procedure. For example, we may want to use a deeper ansatz for the first unitary acting on all the qubits, and shallower ones for smaller subsystems. We thus introduce a different $n'_j$ for each $V_j(\vec{\theta}_j)$ in Eq.~\eqref{eq:local_unitary} and we name \textit{structure} the string `$n_1n_2n_3$'. The latter denotes a learning model consisting of three optimization layers: $V_1(\vec{\theta}_1)$ with  $n_1$ entangling cycles, $V_2(\vec{\theta}_2)$ with  $n_2$ and $V_3(\vec{\theta}_3)$ with  $n_3$, respectively. In the last step of the local optimization procedure, i.e.\ when a single qubit is involved, we always assume a single 3-parameter rotation, see Eq.~\eqref{eq:u3}. A similar notation will be also applied in the following when scaling up to $N>4$ qubits.

\begin{figure}
    \centering
    \includegraphics[width=\columnwidth]{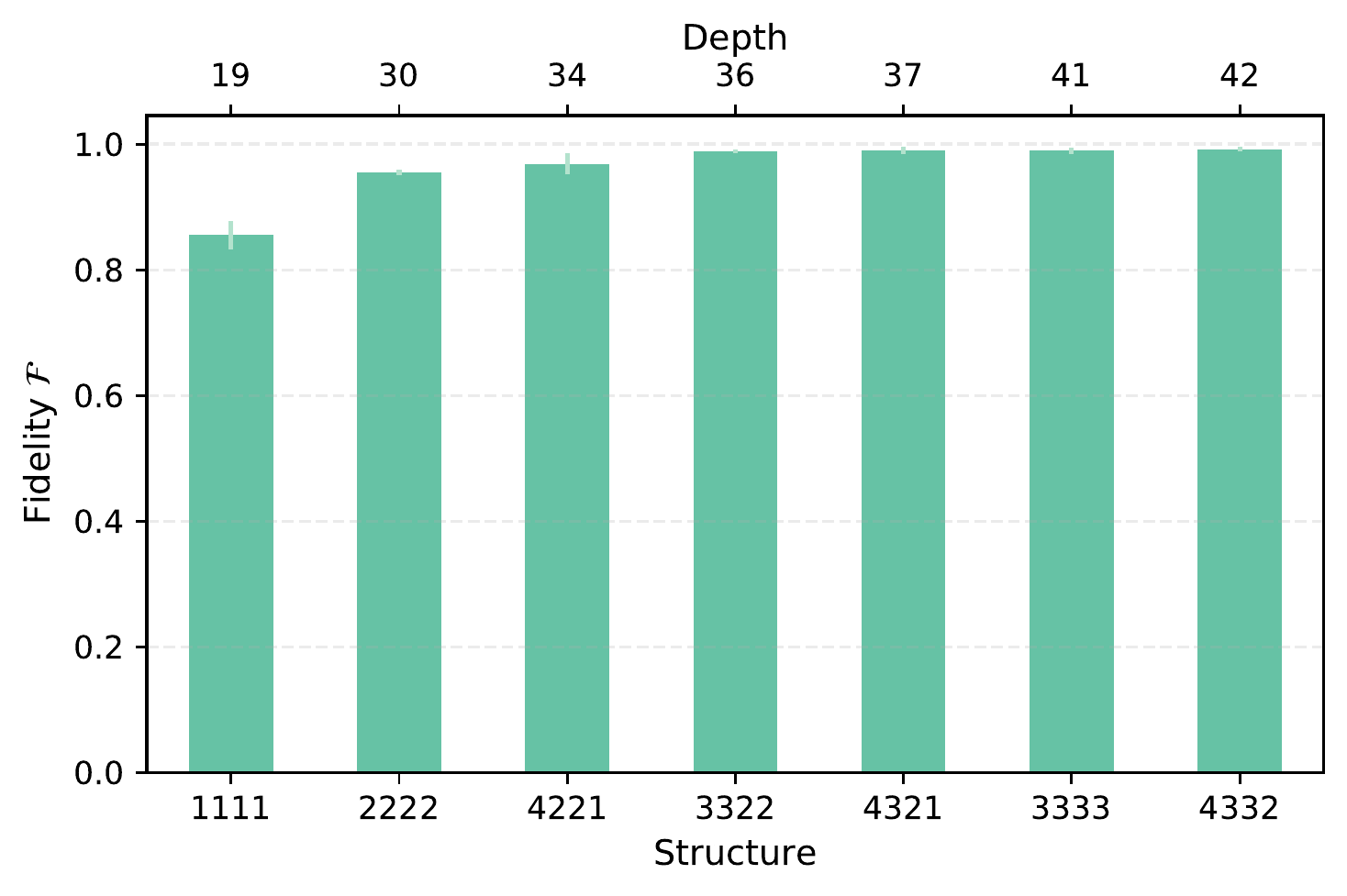}
    \caption{Final fidelities for different structures of the local variational learning model with a nearest neighbour entangler, for the case of $N=5$ qubits. Similarly to the case with $N=4$ qubits portrayed in Figure~\ref{fig:nn_a2a_entangler}, the most depth-efficient structure is the one consisting of constantly decreasing number of cycles.}
    \label{fig:structure_5}
\end{figure}

The effectiveness of different structures is explored in Figure~\ref{fig:nn_a2a_entangler}. We see that, while the all-to-all entangling scheme typically performs better in comparison to the nearest neighbour one, this increase in performance comes at the cost of deeper circuits. Moreover, the stepwise decreasing structure `$321$' for the nearest neighbour entangler proves to be an effective solution to problem, achieving a good final accuracy (above $99\%$) with a low circuit depth. This trend is also confirmed for the higher dimensional case of $N=5$ qubits, which we report in Fig.~\ref{fig:structure_5}. Here, the dimension of the underlying pattern recognition task is increased by extending the original 16-bit weight vector $\vec{w}$ with extra 0s in front of the binary representation $\mathtt{k}_w$. In fact, it can easily be seen that, assuming directly nearest neighbours entangling blocks, the decreasing structure `$4321$'  gives the best performance-depth tradeoff.
Such empirical fact, namely that the most efficient structure is typically the one consisting of decreasing depths, can be heuristically interpreted by recalling again that, in general, the optimization of a function depending on the state of a large number of qubits is a hard training problem~\cite{mcclean_barren_2018}. Although we employ local cost functions, to complete our particular task each variational layer needs to successfully disentangle a single qubit from all the others still present in the register. It is therefore not totally surprising that the optimization carried out in larger subsystems requires more repetitions and parameters (i.e.\ larger $n'_j$) in order to make the ansatz more expressive.

\begin{figure}
    \centering
    \includegraphics[width=\columnwidth]{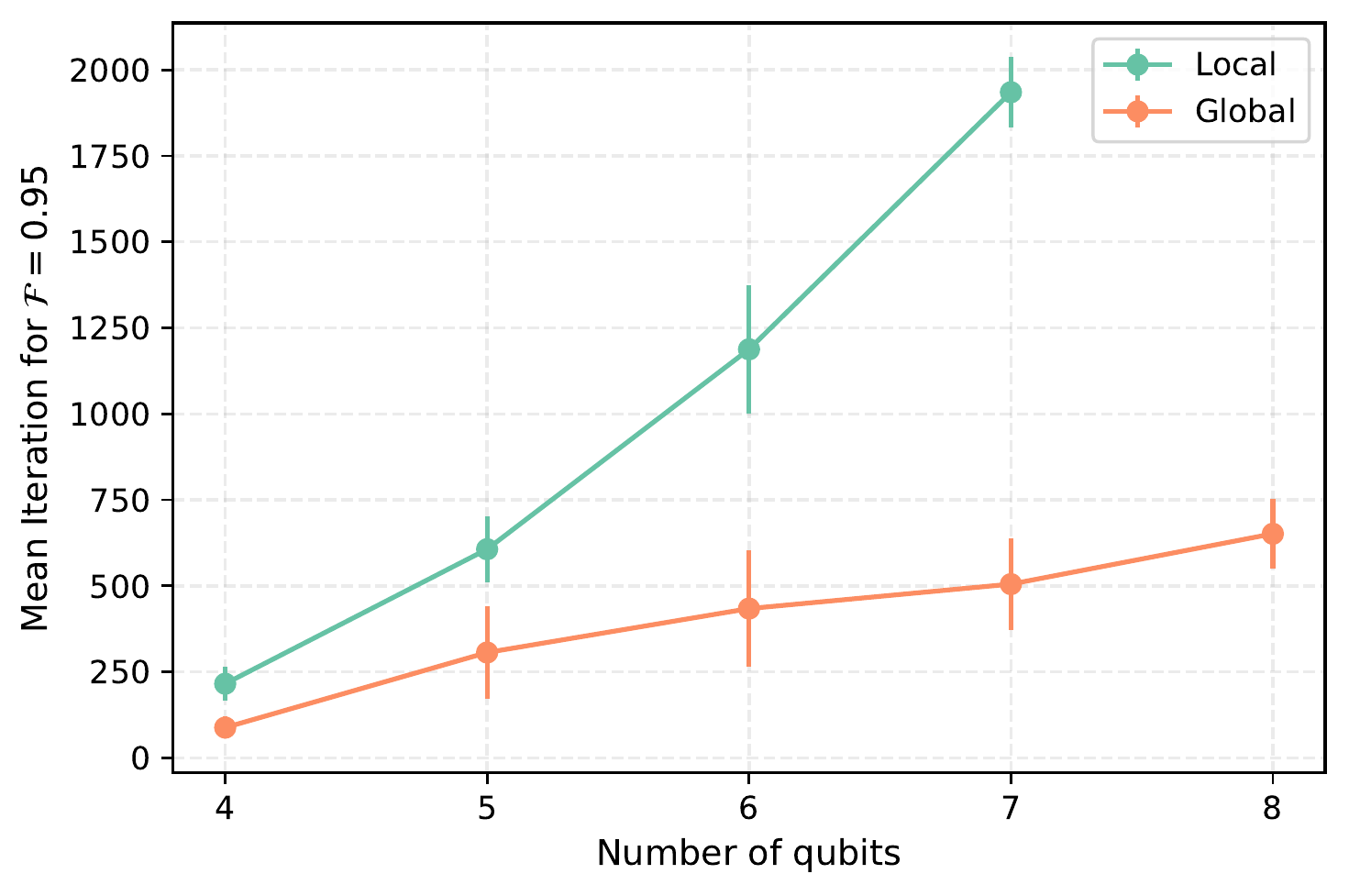}
    \caption{Number of iterations of the classical optimizer to reach a fidelity of $\mathcal{F}=95\%$. Each point in the plot is obtained by running the optimization procedure 10 times and then evaluating the mean and standard deviation (shown as error bars in the plot). All results refer to exact simulations of the quantum circuits in the absence of statistical measurement sampling or device noise, performed with Qiskit \texttt{statevector{\_}simulator}.}
    \label{fig:scalings}
\end{figure}

By assuming that the stepwise decreasing structure remains sufficiently good also for larger numbers of qubits, we studied the optimization landscape of global, Eq.~\eqref{eq:global_cost_f}, and local, Eq.~\eqref{eq:local_cost_f}, cost functions by investigating how the hardness of the training procedure scales with increasing $N$. As commented above for $N = 5$, we keep the same underlying target $\vec{w}$, which we expand by appending extra $0$s in the binary representation. To account for the stochastic nature of the optimization procedure, we run many simulations of the same learning task and report the mean number of iterations needed for the classical optimizer to reach a given target fidelity $\mathcal{F} = 95\%$. Results are shown in Figure~\ref{fig:scalings}. The most significant message is that the use of the aforementioned local cost function seems to require higher classical resources to reach a given target fidelity when the number of qubits increases. 
This actually should not come as a surprise, since the number of parameters to be optimized in the two cases is different. In fact, in the global scenarios there are $N+N\cdot n$ (the first $N$ is due to the initial layer of rotations) to be optimized, while in the local case there are $N+N\cdot n'_1$ for the first layer, $(N-1) + (N-1)\cdot n'_2$ for the second, ...; for a total of 
\begin{equation}
    \#_{\text{local}} = \sum_{q=2}^N q + qn'_q + 3
\end{equation}
where the final $3$ is due to the fact that the last layer always consist of a rotation on the Bloch sphere with three parameters, see Eq.~\eqref{eq:u3}. Using the stepwise decreasing structure, that is $n'_q=q-1$, we eventually obtain $\sum_{q=2}^N q + q(q-1) = \sum_{q=2}^N q^2  \sim O(N^3)$, compared to $\#_{\text{global}} \sim O(N^2)$. Here we are assuming a number of layers $n = N-1$, consistently with the $N=4$ qubits case (see Figure~\ref{fig:global_free}). While in the global case the optimization makes full use of the available parameters to globally optimize the state towards $|1\rangle^{\otimes N}$, the local unitary has to go through multiple disentangling stages, requiring (at least for the cases presented here) more classical iteration steps. At the same time, it would probably be interesting to investigate other examples in which the number of parameters between the two alternative schemes remains fixed, as this would most likely narrow the differences and provide a more direct comparison. 

In agreement with similar investigations~\cite{skolik2020_layerwise}, we can actually conclude that only modest differences between global and local layer-wise optimization approaches are present when dealing with exact simulations (i.e.\ free from statistical and hardware noise) of the quantum circuit. Indeed, both strategies achieve good results and a final fidelity $\mathcal{F}({\vec{\theta}})>99\%$. At the same time, it becomes interesting to investigate how the different approaches behave in the presence of noise, and specifically statistical noise coming from measurements operations. For this reason, we implemented the measurement sampling using the Qiskit \texttt{qasm{\_}simulator} and employed a stochastic gradient descent (SPSA) classical optimization method. Each benchmark circuit is executed $n_{shots}= 1024$ times in order to reconstruct the statistics of the outcomes. Moreover, we repeat the stochastic optimization routine multiple times to analyze the average behaviour of the cost function.
\begin{figure}
    \centering
    \begin{tikzpicture}
    \node[inner sep=0pt] (russell) at (0,0){\includegraphics[width=\columnwidth]{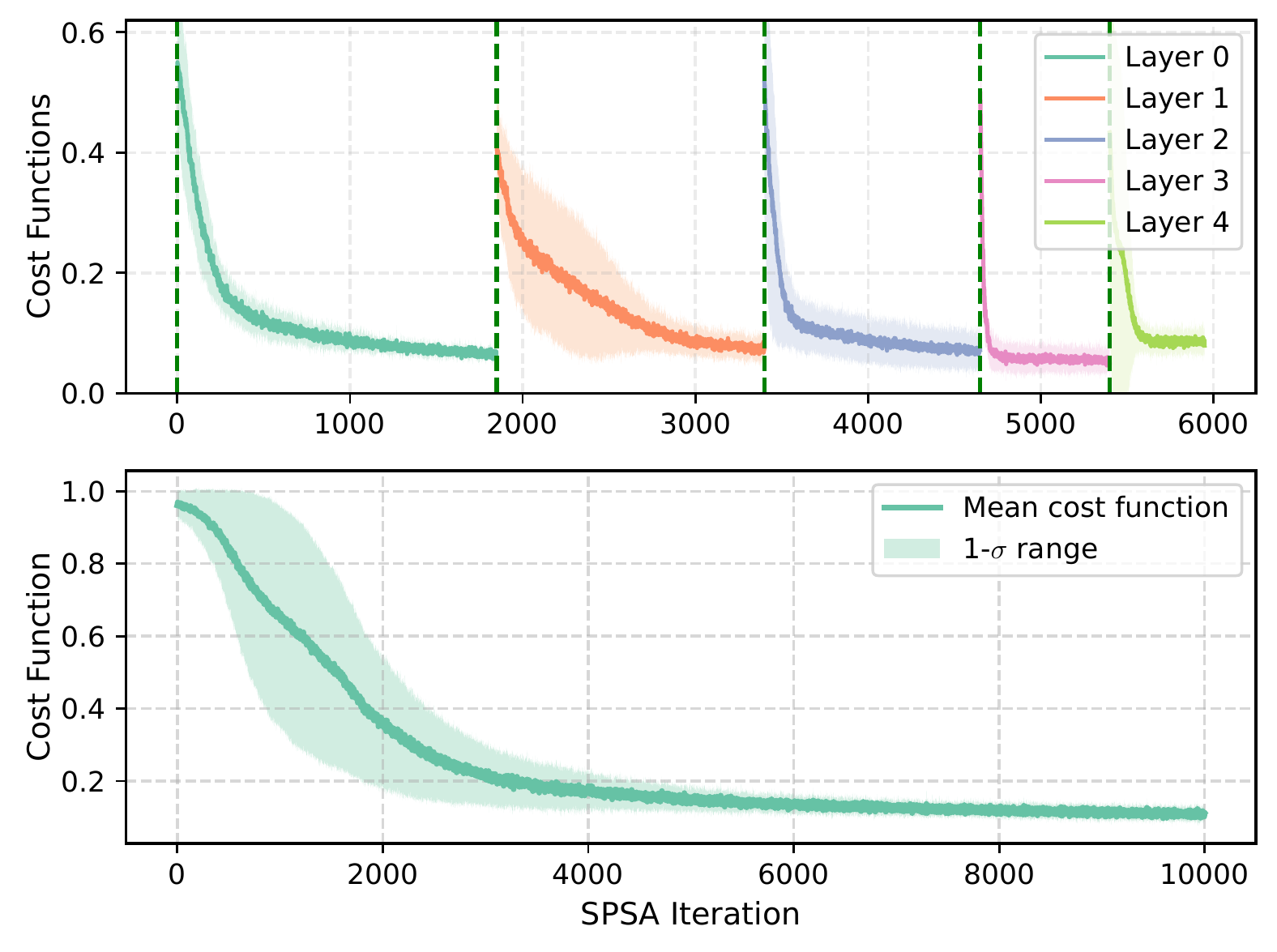}};
    \node[inner sep=0pt] (russell) at (0.1,2.9) {\normalsize{\textbf{a}}};
    \node[inner sep=0pt] (russell) at (0.1,-0.25) {\normalsize{\textbf{b}}};
    \end{tikzpicture}
    \caption{Optimization of cost functions for the local (a) and global (b) case in the presence of measurement noise for $N=5$ qubits. In each figure we plot the mean values averaged on 5 runs of the simulation. The shaded colored areas denote one standard deviation. The number of measurement repetitions in each simulation was 1024. The final fidelity at the end of the training procedure in this case were $\mathcal{F_\text{local}} = 0.87 \pm 0.02$ and $\mathcal{F}_\text{global} = 0.89 \pm 0.02 $. Notice the difference in the horizontal axes bounds. (a) Optimization of the local cost functions $V_j(\vec{\theta}_j)$ (see Eq.~\eqref{eq:local_cost_f}), plotted with different colors for clarity. The vertical dashed lines denotes the end of the optimization of one layer, and the start of the optimization for the following one. (b) Optimization of the global cost function $V(\vec{\theta})$ in Eq.~\eqref{eq:global_cost_f}}. 
    \label{fig:qasm_noise}
\end{figure}
In Figure~\ref{fig:qasm_noise} we show the optimization procedure for the local and global cost functions in the presence of measurement noise, with both of them reaching acceptable and identical final fidelities $\mathcal{F_\text{local}} = 0.87 \pm 0.02$ and $\mathcal{F}_\text{global} = 0.89 \pm 0.02 $. Notice that for the local case (Figure~\ref{fig:qasm_noise}(a)) each colored line indicates the optimization of a $V_j(\vec{\theta}_j)$ from Eq.~\eqref{eq:local_cost_f}. We observe that the training for the local model generally requires fewer iterations, with an effective optimization of each single layer. On the contrary, in the presence of measurement noise the global variational training struggles to find a good direction for the optimization and eventually follows a slowly decreasing path to the minimum. These findings look to be in agreement, e.g., with results from Refs.~\cite{skolik2020_layerwise, cerezo2020costfunctiondependent}: with the introduction of statistical shot noise, the performances of the global model are heavily affected, while the local approach proves to be more resilient and capable of finding a good gradient direction in the parameters space~\cite{cerezo2020costfunctiondependent}. In all these simulations, the parameters in the global unitary and in the first layer of the local unitary were initialized with a random distribution in $[0, 2\pi)$. All subsequent layers in the local model were initialized with all parameters set to zero in order to allow for smooth transitions from one optimization layer to the following. This strategy was actually suggested as a possible way to mitigate the occurence Barren plateaus~\cite{skolik2020_layerwise, Grant2019initialization}. 

\begin{figure}[!t]
\includegraphics[width=\columnwidth]{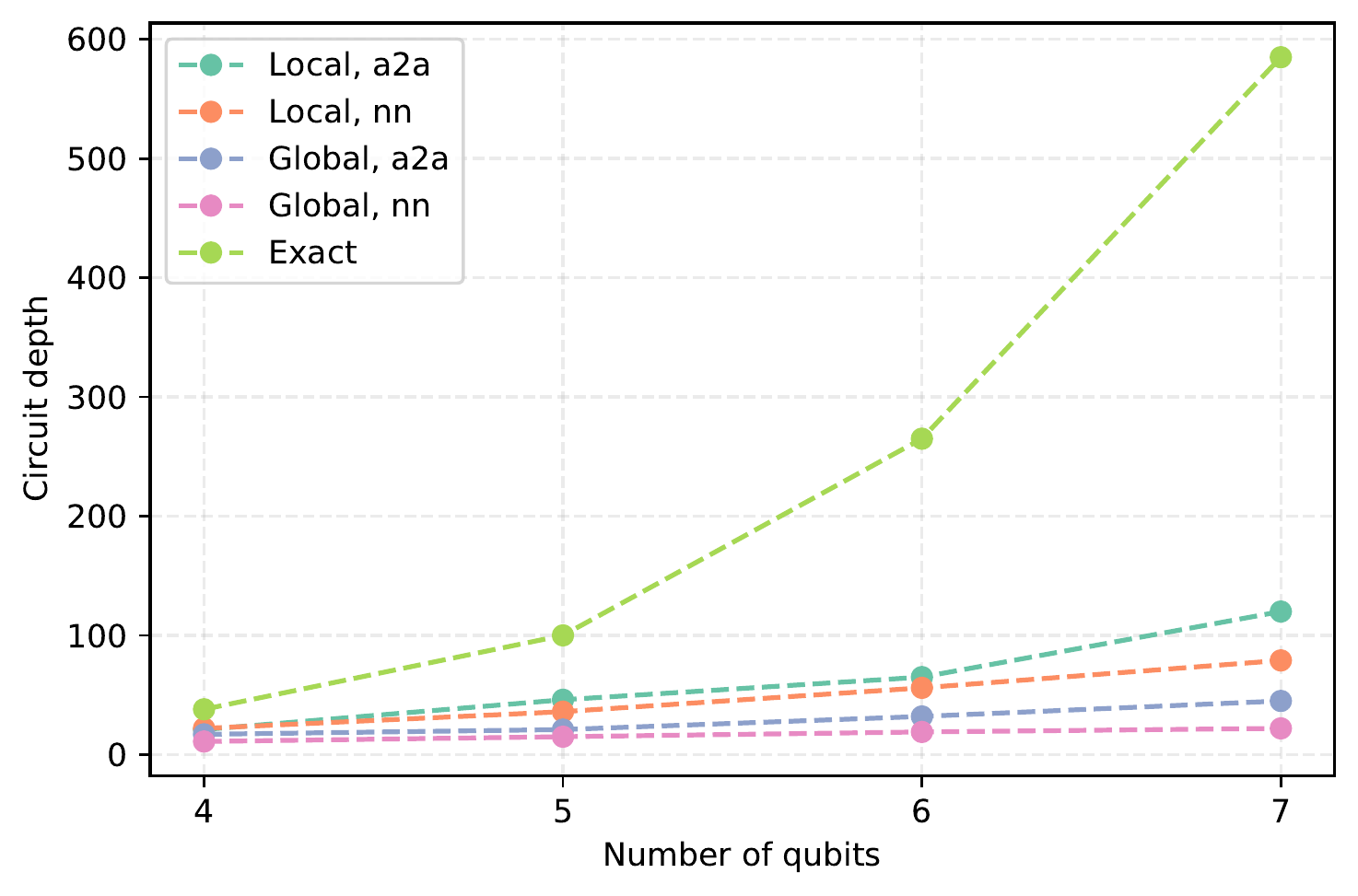}
\caption{Scaling of circuit depth for the implementation of $\mathrm{U}_w$ computed with Qiskit. The labels \textit{locals} and \textit{global} refer to the local and global variational approaches, while \textit{a2a} and \textit{nn} refer to the all-to-all and nearest-neighbour entangling schemes respectively. The number of ansatz cycles used for both the global ($n$)  and local/qubit-by-qubit ($n'$) variational constructions and for each entangling sctructure are increased with the number of qubits up to the minimum value guaranteeing a fidelity of the approximations above $98\%$.
}
\label{fig:scaling_depth}
\end{figure}

We conclude the scaling analysis by reporting in Fig.~\ref{fig:scaling_depth} a summary of the quantum circuit depths required to implement the target unitary transformation with different strategies and for increasing sizes of the qubit register up to $N = 7$. As it can be seen, all the variational approaches scale much better when compared to the exact implementation of the target $\mathrm{U}_w$, with the global ones requiring shallower depths in the specific case. In addition, we recall that the use of an all-to-all entangling scheme requires longer circuits due to the implementation of all the CNOTs, but generally needs less ansatz cycles (see Figure~\ref{fig:nn_a2a_entangler}). At last, while the global procedures seem to provide a better alternative compared to local ones in terms of circuit depth, they might be more prone to suffering from classical optimization issues~\cite{skolik2020_layerwise, mcclean_barren_2018} when trained and executed on real hardware, as suggested by the data reported in Fig.~\ref{fig:qasm_noise}. The overall promising results confirm the significant advantage brought by variational strategies compared to the exponential increase of complexity required by the exact formulation of the algorithm.

\section{Conclusions}

In this work, we reviewed an exact model for the implementation of artificial neurons on a quantum processor and we introduced variational training methods for efficiently handling the manipulation of classical and quantum input data. Through extensive numerical analysis, we compared the effectiveness of different circuit structures and learning strategies, highlighting potential benefits brought by hardware-compatible entangling operations and by layerwise training routines. Our work suggests that quantum unsampling techniques represent a useful resource, upon input of quantum training sets, to be integrated in quantum machine learning applications. From a theoretical perspective, our proposed procedure allows for an explicit and direct quantification of possible quantum computational advantages for classification tasks. It is also worth pointing out that such a scheme remains fully compatible with recently introduced architectures for quantum feed-forward neural networks~\cite{tacchino_quantum_2019}, which are needed in general to deploy e.g.\ complex convolutional filters. Moreover, although the interpretation of quantum hypergraph states as memory-efficient carriers of classical information guarantees an optimal use of the available dimension of a $N$-qubit Hilbert space, the variational techniques introduced here can in principle be used to learn different encoding schemes designed, e.g., to include continuous-valued features or to improve the separability of the data to be classified~\cite{buhrman_quantum_2001,havlicek_supervised_2019,schuld_quantum_2019}. In all envisioned applications, our proposed protocols are intended as an effective method for the analysis of quantum states as provided, e.g., by external devices or sensors, while it is worth stressing that the general problem of efficiently loading classical data into quantum registers still stands open. Finally, on a more practical level, a successful implementation on near-term quantum hardware of the variational learning algorithm introduced in this work will necessarily rely on a deeper analysis of the impact of realistic noise effects both on the training procedure and on the final optimized circuit. In particular, we anticipate that the reduced circuit depth produced via the proposed method could critically lessen the quality requirements for quantum hardware, eventually leading to meaningful implementation of quantum neural networks within the near-term regime. 

\section*{Acknowledgments}
A preliminary version of this work was presented at the 2020 IEEE International Conference on Quantum Computing and Engineering\cite{tacchino_IEEE_2020}. We acknowledge support from SNF grant 200021\_179312. IBM, the IBM logo, and ibm.com are trademarks of International Business Machines Corp., registered in many jurisdictions worldwide. Other product and service names might be trademarks of IBM or other companies. The current list of IBM trademarks is available at \url{https://www.ibm.com/legal/copytrade}.


\end{document}